\documentclass[prl,twocolumn,preprintnumbers,amsmath,amssymb,showpacs]{revtex4}
\usepackage{amsfonts}
\usepackage{bbm}
\usepackage{mathrsfs}
\usepackage{amsmath}
\usepackage{color}
\usepackage{graphicx}
\usepackage{epstopdf}
\usepackage{dcolumn}
\usepackage{bm}
\usepackage{longtable}
\usepackage{graphics}
\usepackage{amssymb}
\usepackage{xspace}
\usepackage{epsfig}
\usepackage{subfigure}
\usepackage{dcolumn}
\usepackage{multirow}
\makeatletter

\newcommand{\Rmnum}[1]{\expandafter\@slowromancap\romannumeral #1@}
\makeatother

\begin{document}
\title{Surface States of Topological Insulators}
\author{Fan Zhang}\email{zhf@sas.upenn.edu}
\author{C. L. Kane}
\author{E. J. Mele}
\affiliation{Department of Physics and Astronomy, University of Pennsylvania, Philadelphia, PA 19104, USA}
\begin{abstract}
We develop an effective bulk model with a topological boundary condition to study the surface states of topological insulators.
We find that the Dirac point energy, the band curvature and the spin texture of surface states are crystal face-dependent.
For a given face on a sphere, the Dirac point energy is determined by the bulk physics that breaks p-h symmetry in the surface normal direction and is tunable by surface potentials that preserve $\mathcal{T}$ symmetry. Constant energy contours near the Dirac point are ellipses with spin textures that are helical on the S/N pole, collapsed to one dimension on any side face, and tilted out-of-plane otherwise.
Our findings identify a route to engineering the Dirac point physics on the surfaces of real materials.
\end{abstract}
\date{\today}
\maketitle

\indent{\em Introduction.}---The discovery\cite{Z2,Fu_Kane_Mele,Moore_Balents,Roy,Kane_RMP,Zhang_RMP} of topological insulators (TI's)
and the synthesis\cite{BiSb,BiSe,BiTe,Zhang_DFT} of three dimensional materials that realize their physics has
opened up a new field in solid state physics. Particular interest has focused on the TI surface states with point degeneracies
that are topologically protected by time reversal symmetry ($\mathcal{T}$) when a trivial insulator with topological index\cite{Z2,Fu_Kane_Mele}
$\mathcal{Z}_2=1$ is joined to a TI with $\mathcal{Z}_2=-1$. On the cleavage surface of ${\rm Bi_2Se_3}$, a TI with a single
Dirac cone has been identified\cite{BiSe,BiTe,STM1,STM2,Xue_STM,STM3,STM_warping} by angle resolved photoemission (ARPES) and scanning tunneling microscope (STM) experiments. The
surface states form a spin-momentum locked helical metal with conduction and valence bands exhibiting opposite
helicities near the Dirac point, and develop hexagonal warping\cite{BiTe,STM_warping,Fu_warping,Liu} on top of the circular constant energy contour away from the Dirac point.

Interest in the topological surface bands has led to the development of various bulk continuum models with the surface termination
described by ad-hoc fixed node \cite{Zhang_DFT,Liu,Sudbo,Shen,Niu} or by  ``natural" boundary conditions \cite{Medhi}, where the existence of a
topological surface state and its Dirac point energy are treated as inputs to the theories. The choice of boundary condition has profound consequences for the spatial form of the surface state wavefunctions and their interactions with external fields and absorbed species.
In this work we derive the appropriate {\em topological boundary condition} (TBC) at the surface
from the matching of bulk waves to evanescent vacuum states. As expected a topologically protected Dirac point
occurs in this model but its energy position and band curvature are both crystal face-dependent owing to bulk terms that break
particle-hole (p-h) symmetry. Furthermore, the interaction with surface-localized potentials that preserve $\mathcal{T}$ symmetry shifts
the energy position of the Dirac degeneracy and provides a robust degree of freedom for tuning the surface state spectrum into a convenient energy range to exploit their topological properties.
The associated spin texture is determined by the bulk symmetries and depends on the crystal face angle.
Unlike the helical metal on a cleavage surface, the spin texture near Dirac point is compressed to a single direction on a side
face, and is tilted out-of-plane otherwise.

We start from a description of the low-energy model of ${\rm Bi_2Se_3}$, which applies generally to other TI's with the
same crystal structure with space group $R3\bar{m}$. At the origin of ${\rm Bi_2Se_3}$ Brillouin zone $\Gamma$, the effective
Hilbert space near the bulk gap is spanned by states with angular momentum $m_{\rm j}=\pm \frac{1}{2}$ and parity
$\mathcal{P}=\pm 1$. Because of the spin-orbit coupling (SOC), $|p_{\rm z}\uparrow\rangle$ and $|p_{\rm z}\downarrow\rangle$
states are mixed with $|p_{+}\downarrow\rangle$ and $|p_{-}\uparrow\rangle$ states, respectively. Since the crystal-field
splitting is much larger than the SOC, $p_{\rm z}$ orbitals dominate and $m_{\rm j}$ pseudospin is proportional to the electron
spin. The $\mathcal{P}=\pm 1$ hybridized states can be labelled approximately by $|+\rangle$ state from ${\rm Bi}$ atoms while
$|-\rangle$ from ${\rm Se}$, due to the large energy difference between $4p$ (${\rm Se}$) and $6p$ (${\rm Bi}$) principal
quantum levels.

Besides $\mathcal{T}$ and the parity inversion ($\mathcal{P}$) symmetries, ${\rm Bi_2Se_3}$ crystal
structure has threefold rotational ($\mathcal{C}_3$) symmetry along the $\hat{z}$ perpendicular to the quintuple layers, and
twofold rotational ($\mathcal{C}_2$) symmetry along $\Gamma M$ direction. By
convention we choose the parity operator $\mathcal{P}=\tau_{\rm z}$ and the time reversal operator $\mathcal{T}=iK\sigma_{\rm
y}$ where $K$ is the complex conjugate operation. Therefore, to quadratic order in $k$ the ${\bm k}\cdot{\bm p}$ bulk
Hamiltonian that preserves the above four symmetries has a unique form
\begin{eqnarray}
\label{eqn:H}
\mathcal{H}&=&(c_0+c_{\rm z}k_{\rm z}^2+c_{\shortparallel}k_{\shortparallel}^2)
           +(-m_0+m_{\rm z}k_{\rm z}^2+m_{\shortparallel}k_{\shortparallel}^2)\tau_{\rm z}\nonumber\\
           &+&v_{\rm z}k_{\rm z}\tau_{\rm y}+v_{\shortparallel}(k_{\rm y}\sigma_{\rm x}-k_{\rm x}\sigma_{\rm y})\tau_{\rm x}\,,
\end{eqnarray}
where we assume $m_{\rm z}, m_{\shortparallel}, v_{\rm z}, v_{\shortparallel}>0$ and $\hbar=1$ hereafter. The first parentheses is a scalar
term that preserves $\mathcal{T}$ and $\mathcal{P}$ symmetries but breaks the p-h symmetry. The quadratic
scalar terms intrinsically give rise to the curvature of Dirac surface bands and the rigid shift of Dirac point from the middle
of the bulk gap, as we will demonstrate in the following. Eq.(\ref{eqn:H}) has both $\mathcal{T}$ and $\mathcal{P}$ symmetries
with a topological index $\mathcal{Z}_2={\rm sgn}(-m_0)$.

\indent{\em Topological boundary condition.}---
To investigate the surface states, we use a TBC in which the mass term changes sign across the
surface. On the TI side, $m_0=m$ is positive and half of the bulk gap at ${\bm k}=0$, and $c_0=0$ to define the middle of the
${\bm k}=0$ gap as energy zero. On the vacuum side, {\em i.e.}, a trivial insulator with an infinite gap $m_0=-M$ where $M\rightarrow +\infty$, both $c_{\rm z}$ and $c_{\shortparallel}$ vanish since the vacuum has p-h symmetry around $c_0$.
\begin{figure}[t]
\centering{ \scalebox{0.55} {\includegraphics*{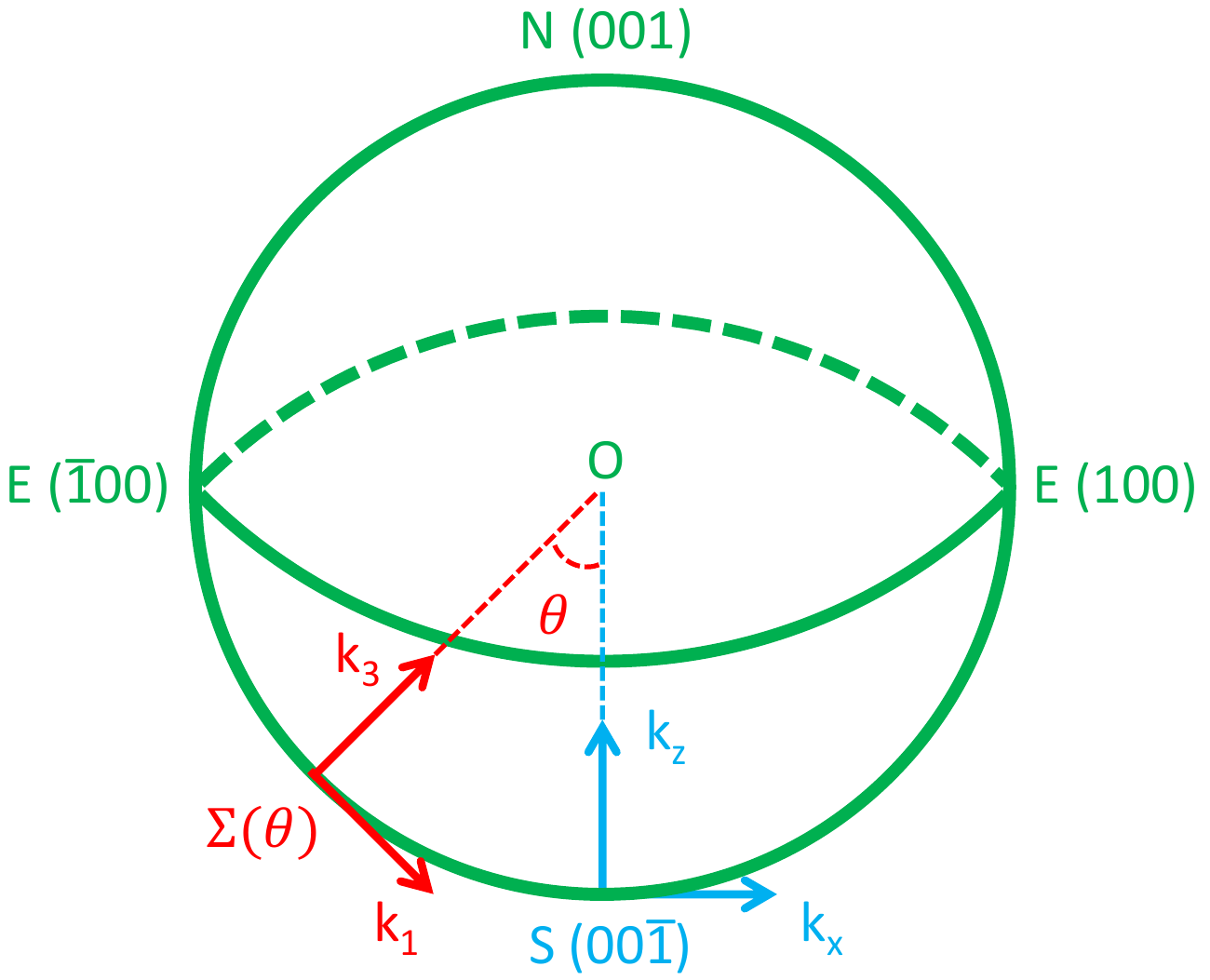}}}
\caption{\label{fig:frame} {(Color online) The definition of an arbitrary face of a $Bi_2Se_3$-type topological insulator. The south (S) pole denotes the $(00\bar{1})$ surface parallel to the quintuple layers and the equator represents the side faces perpendicular to the quintuple layers. For any face $\Sigma(\theta)$, ${\hat{k}}_3\perp\Sigma$ and ${k}_2={k}_{\rm y}$.}}
\end{figure}

We first focus on the one dimensional quantum mechanics of $(00\bar{1})$ surface turning off the quadratic terms in
Eq.(\ref{eqn:H}). Notice that the two spin flavors are decoupled for this eigenvalue problem. The spin $\uparrow$ states in
$\{\tau_{\rm z},\sigma\}$ representation are determined by the following coupled Dirac equations
\begin{eqnarray}\label{eqn:D}
\left(\begin{array}{cc} -m(z)-E & -v_{\rm z}\partial_{\rm z}\\ v_{\rm z}\partial_{\rm z} & m(z)-E\\ \end{array}\right)
\left(\begin{array}{c} \psi_1 \\ \psi_2 \\ \end{array}\right) =0\,.
\end{eqnarray}
The wave function $\psi_{\uparrow}=(\psi_1,\psi_2)'$ is continuous across the surface, integrating Eq.(\ref{eqn:D}) over the vicinity of the surface. This continuity condition leads to a nontrivial solution which is isolated in the middle of the bulk gap and is localized on the surface. This midgap ($E=0$) state has a simple and elegant exact solution,
\begin{eqnarray}
\label{eqn:wf}
\psi_{k_{\shortparallel},\uparrow}(x,y,z)&=&\frac{1}{A} e^{i(k_{\rm x}x+k_{\rm y}y)}\phi(z)
\left(\begin{array}{c} 1 \\ 0 \\ \end{array}\right)_{\rm \sigma}
\otimes\left(\begin{array}{c} 1 \\ 0 \\ \end{array}\right)_{\rm \tau_x}\,,\nonumber\\
\phi(z)&=&\begin{cases}
 e^{-\kappa z}, & z>0 \quad(\text{TI}) \\
 e^{\kappa_0 z}, & z<0 \quad(\text{Vac})
\end{cases}\,,
\end{eqnarray}
where $\kappa=m/v_{\rm z}$, $\kappa_0=M/v_{\rm z}$ and $A$ is a normalization factor. $\phi(z)$ is evanescent on both sides of the surface where the mass $m_0$ changes sign, analogous to Jackiw and Rebbi solution\cite{Jackiw_Rebbi} of a two-band Dirac model. The spin $\downarrow$ solution can be obtained by $\psi_{\downarrow}=\mathcal{T}\psi_{\uparrow}$. This isolated midgap state at $k_{\shortparallel}=0$ is identified as the Dirac point and at finite $k_{\shortparallel}$ spreads into a perfect Dirac cone-the ideal topologically protected surface bands.

\begin{figure}[t]
\centering{\scalebox{0.54}{\includegraphics*{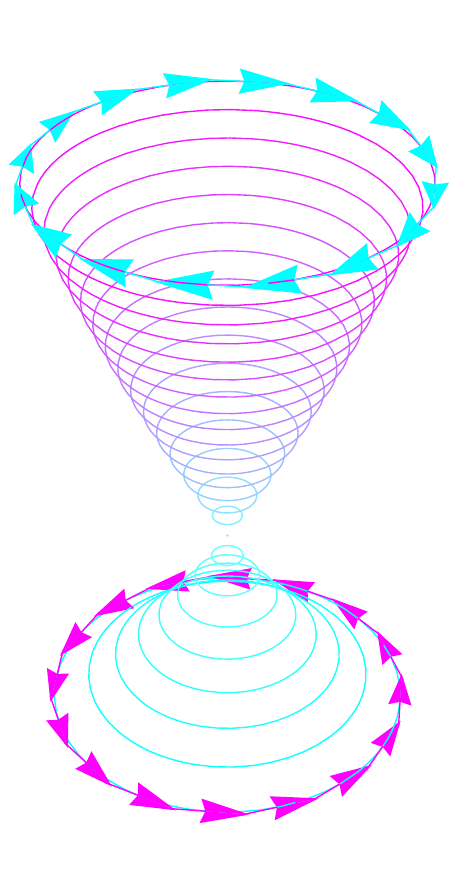}}\;\;\scalebox{0.54}{\includegraphics*{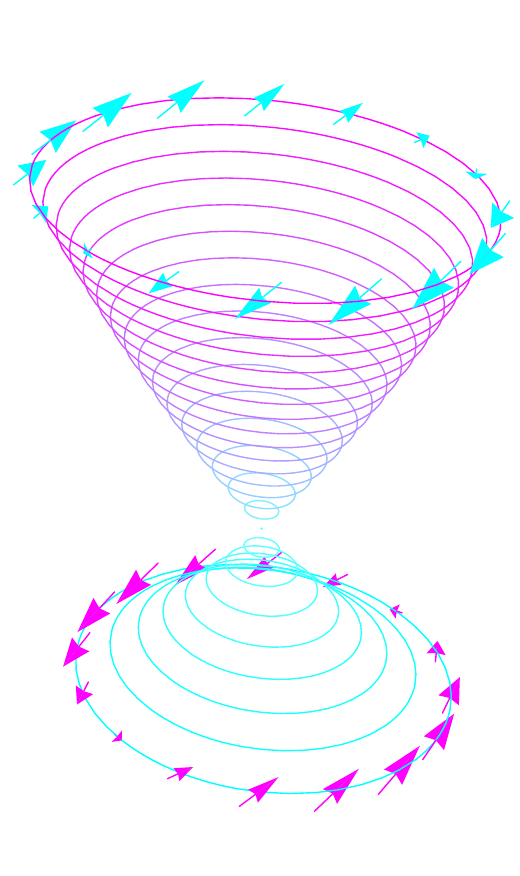}}\scalebox{0.54}{\includegraphics*{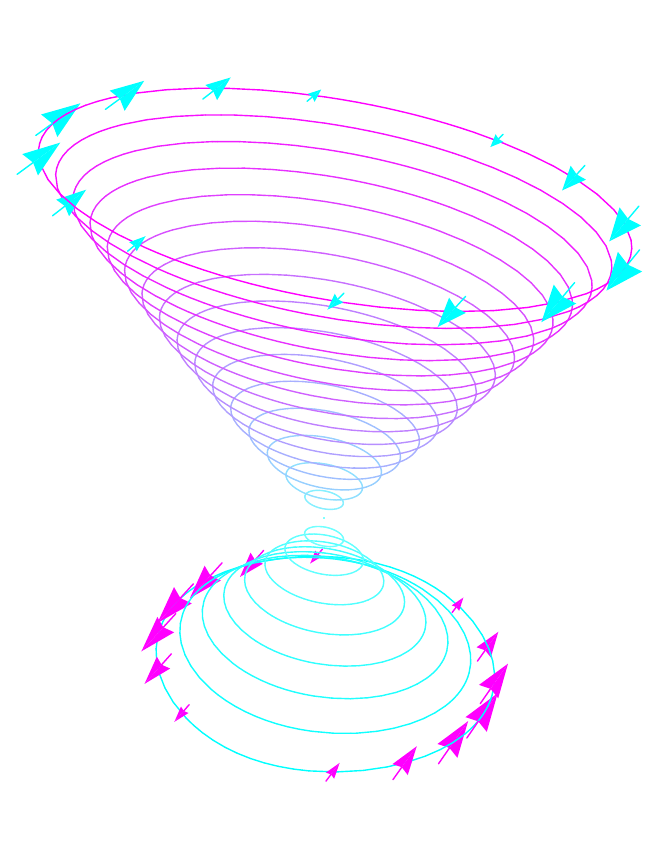}}
\scalebox{0.50}{\includegraphics*{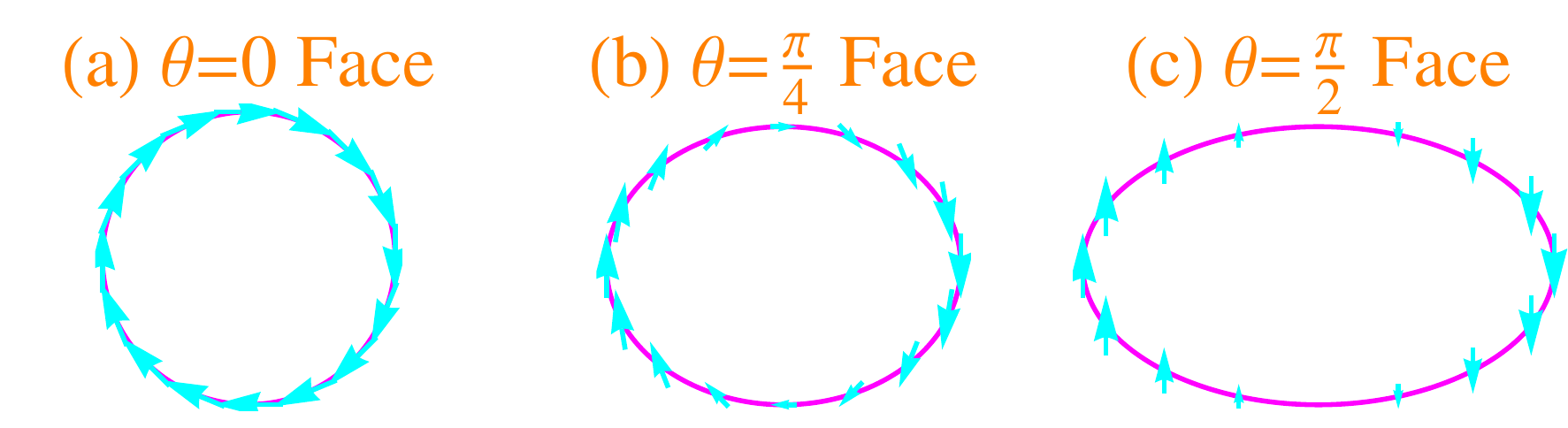}}}
\caption{\label{fig:spintexture} {(Color online) Dirac cones and spin textures of surface states on faces with (a) $\theta=0$, (b) $\theta={\pi}/{4}$ and (c) $\theta={\pi}/{2}$. The equal-energy contours are from $-80$ meV to $160$ meV with $10$ meV increment relative to the Dirac point. Surface band curvatures are taken into account. All three panels are in the same scale and the parameters are adopted from DFT results. The lower panels only show the spin textures at $160$ meV in the $k_1-k_{\rm y}$ planes where $k_{\rm y}$ is the vertical axis.}}
\end{figure}

\indent{\em Surface state spin texture.}---
On the $(00\bar{1})$ surface, the midgap solution of the boundary problem at $k_{\shortparallel}=0$ is determined by the operators ${\bm \tau}$ and is free under any rotation of the operators ${\bm \sigma}$. This ${\bm \tau}\otimes{\bm \sigma}$ algebra of the boundary problem can be generalized to any surface with ${\bm \tau}$ replaced by ${\bm S}_1$ and ${\bm \sigma}$ by ${\bm S}_2$. For an arbitrary crystal face $\Sigma(\theta)$ defined in Fig.\ref{fig:frame}, the algebraic structure is
\begin{eqnarray}
\label{eqn:structure}
{\bm S}_1&=\{\alpha\tau_{\rm x}+\beta\sigma_{\rm y}\tau_{\rm y},\alpha\tau_{\rm y}-\beta\sigma_{\rm y}\tau_{\rm x},\tau_{\rm z}\}\,,&\nonumber\\
{\bm S}_2&=\{\alpha\sigma_{\rm x}-\beta\sigma_{\rm z}\tau_{\rm z},\sigma_{\rm y},\alpha\sigma_{\rm z}+\beta\sigma_{\rm x}\tau_{\rm z}\}\,,&
\end{eqnarray}
where $v_3=\sqrt{(v_{\rm z}\cos\theta)^2+(v_{\shortparallel}\sin\theta)^2}$, $\alpha=v_{\rm z}\cos\theta/v_3$ and $\beta=v_{\shortparallel}\sin\theta/v_3$. These new pseudospins satisfy $[S_{\rm a}^{\rm i},S_{\rm b}^{\rm j}]=2i\delta_{\rm ab}\epsilon^{\rm ijk}S_{\rm a}^{\rm k}$. Rewritten in this new pseudospin basis, Eq.(\ref{eqn:H}) reads
\begin{eqnarray}
\label{eqn:newH}
\mathcal{H}=&-& m_0\, S_1^{\rm z}+(v_3 k_3+v_0 k_1)\, S_1^{\rm y}\nonumber\\
&+&(v_{\shortparallel}k_{\rm y}\,S_2^{\rm x}-v_1 k_1\, S_2^{\rm y})\,S_1^{\rm x}\,,
\end{eqnarray}
where $v_0=(v_{\shortparallel}^2-v_{\rm z}^2)\sin\theta\cos\theta/v_3$ and $v_1=v_{\rm z}v_{\shortparallel}/v_3$. By matching the eigensystems of TI and vacuum sides, we obtain the $\Sigma(\theta)$ surface states similar to Eq.(\ref{eqn:wf}) where ${\bm \tau}$ and ${\bm \sigma}$ are replaced by ${\bm S}_1$ and ${\bm S}_2$, respectively. Note that $\kappa=m/v_3+iv_0k_1/v_3$ in general. $k_1$ coupling to $S_1^{\rm y}$ neither influences the energy spectrum nor the spin texture, only making the evanescent states oscillate and giving a phase accumulation along $\hat{k}_1$ away from Dirac point. Therefore, ignoring the quadratic corrections, we first derive the effective surface state Hamiltonian for an arbitrary face $\Sigma(\theta)$ to the linear order,
\begin{eqnarray}
\mathcal{H}^{(1)}(\theta)=v_{\shortparallel}k_{\rm y}\,S_2^{\rm x}-v_1 k_1\,S_2^{\rm y}\,,
\end{eqnarray}
from which we can further explicitly demonstrate the surface state spin texture on $\Sigma(\theta)$:
\begin{eqnarray}
\label{eqn:spin}
\langle\sigma_{\rm x}\rangle_{\rm \theta}&=&\pm\frac{v_{\rm z}v_{\shortparallel}k_{\rm y}\cos\theta}{v_3\sqrt{v_1^2k_1^2+v_{\shortparallel}^2k_{\rm y}^2}}\,,\nonumber\\
\langle\sigma_{\rm y}\rangle_{\rm \theta}&=&\pm\frac{-v_{\rm z}v_{\shortparallel}k_1}{v_3\sqrt{v_1^2k_1^2+v_{\shortparallel}^2k_{\rm y}^2}}\,,\nonumber\\
\langle\sigma_{\rm z}\rangle_{\rm \theta}&=&0\,,
\end{eqnarray}
where $+\,\,(-)$ denotes the conduction (valence) band. The electron real spin $\langle {\bm s}\rangle$ is proportional to but always smaller than  $\langle{\bm \sigma}\rangle$ due to the SOC\cite{Louie_DFT}. In the local coordinates, $\langle\sigma_{1}\rangle_{\rm \theta}=\langle\sigma_{\rm x}\rangle_{\rm \theta}\cos\theta$ and $\langle\sigma_{3}\rangle_{\rm \theta}=\langle\sigma_{\rm x}\rangle_{\rm \theta}\sin\theta$. Eq.(\ref{eqn:spin}) indicates that the surface state spin texture is rather different from face to face while its pseudospin ($S_2$) has a universal structure on the elliptic constant energy contour near the Dirac point. Clearly, there is no spherical symmetry to guarantee that the spin and orbital structures are the same anywhere on a TI sphere. The crystal face-dependent spin texture is helical on the south and north poles, is compressed to a single dimension along the equator, and is tilted out-of-plane otherwise.  The surface state anisotropy and spin textures for different faces are compared in Fig.\ref{fig:spintexture}.

The surface state spin textures on the poles and the equator of a TI sphere can be understood by symmetries. For $(\bar{1}00)$ face, the $\mathcal{C}_2$ symmetry along $\Gamma M$ ($\hat{x}$) direction forbids $\sigma_{\rm x}$ coupling to any in-plane momentum. In linear order $\mathcal{C}_3$ symmetry upgrades to continuous rotational symmetry, consequently, the surface normal spin is zero along the TI equator. On the two poles, the mirror symmetry and the $\mathcal{C}_3$ symmetry insure that the spin-momentum locking into the form of $k_{\rm y}\sigma_{\rm x}-k_{\rm x}\sigma_{\rm y}$. $\sigma_{\rm z}$ decouples to any momentum in the linear order, taking into account all the four symmetries.

It's interesting to point out that the surface band is the positive eigenstate of $S_{1}^{\rm x}$ and the chiral counterpart is separated and localized on the opposite face. Thus the surface state Hilbert space is reduced by {\em half} and this pseudospin polarity (or chirality) blocks the back scattering on the surface as a result of interplay between $\mathcal{T}$ symmetry and band inversion physics with TBC.

\indent{\em Dirac point energy and surface potentials.}---
The absence of spherical symmetry in the bulk requires that the Dirac point has different energies on different crystal faces. When the quadratic mass terms and p-h symmetry breaking terms in Eq.(\ref{eqn:H}) are turned on, the surface state wave function and spectrum are changed. In the coupled Schr\"{o}dinger-type Eq.(\ref{eqn:H}), the components of wave function and their slopes are all continuous across the boundary. We find that the effective Hamiltonian for face $\Sigma(\theta)$ have two important\cite{ft1} corrections:
\begin{eqnarray}
\label{eqn:curvature}
\mathcal{H}^{(2)}(\theta)&=&c_{\shortparallel}k_{\rm y}^2+(c_{\rm z}\sin^2\theta+c_{\shortparallel}\cos^2\theta)k_{1}^2\,,\\
\label{eqn:DP}
\mathcal{H}^{\rm DP}(\theta)&=&\frac{c_{\rm z}\cos^2\theta+c_{\shortparallel}\sin^2\theta}{m_{\rm z}\cos^2\theta+m_{\shortparallel}\sin^2\theta}\cdot m\,.
\end{eqnarray}
The breaking p-h symmetry terms give the surface Dirac cone a parabolic curvature described by Eq.(\ref{eqn:curvature}) and shifts the Dirac point from the midgap to a nonzero energy given by Eq.(\ref{eqn:DP}).
These two effects help to explain the origin of nonzero Dirac point energies and nonlinear Dirac cones of cleavage surface states observed in ARPES experiments\cite{BiSe,BiTe,Cui,Xue,Ando_2}.

Although the surface state solution obtained by Eq.(\ref{eqn:D}) with TBC remains topologically stable, it is essential to understand how localized surface potentials influence the surface spectra and the associated wave functions as well. We focus on potentials that preserve $\mathcal{T}$ symmetry. Among these six types: $I$, $\tau_{\rm z}$, $\tau_{\rm x}$ and $\vec{\sigma}\tau_{\rm y}$ list in Table \ref{table:one}, $\tau_{\rm x}$ and $\vec{\sigma}\tau_{\rm y}$ break $\mathcal{P}$ symmetry. It turns out that the same potential could play different roles on different crystal faces, as shown in Table \ref{table:one}.

\begin{table}[b]
\caption{Summary of the influence of $\mathcal{T}$ symmetry allowed momentum-independent surface potentials $\Delta\delta(r_3)\cdot2v_{3}/m$ on the inversion symmetry, and the wave function continuity and the Dirac point energy of surface states. This $\Delta $ represents different surface potentials and their corresponding energy scales: $\Delta_0I$, $\Delta_{\rm m}\tau_{\rm z}$, $\Delta_{\rm E}S_1^{\rm x}$ and $\Delta_{\rm n}S_2^{\rm n}S_1^{\rm y}$. For the $\sigma_{\rm n}\tau_{\rm y}$ column, only the results for $S_2^{\rm n}=1$ state is shown and their complex conjugates represent the results for $S_2^{\rm n}=-1$ state.}
\newcommand\T{\rule{0pt}{4ex}}
\newcommand\B{\rule[-1.7ex]{0pt}{0pt}}
\centering
\addtolength{\tabcolsep}{-1.2pt}
\begin{tabular}{ c || c | c | c | c }
\hline\hline
\;$\Sigma(0)$\; & ${I}$ & $\tau_{\rm z}$ & $\tau_{\rm x}$ & $\sigma_{\rm n}\tau_{\rm y}$ \T\\[3pt]
\hline
\;$\Sigma(\theta)$\; & ${I}$ & $\tau_{\rm z}$ &
$S_1^{\rm x}$ &
$S_2^{\rm n}S_1^{\rm y}$ \T\\[3pt]
\hline
$\mathcal{P}$ & $+$ & $+$ & $-$ & $-$ \T \\[3pt]
\hline
$\delta \mathcal{E}_{\rm DP}$  &  $\frac{4m^2\Delta_0(m^2-\Delta_0^2)}{(m^2-\Delta_0^2)^2+4m^2\Delta_0^2}$ & $0$ &
$\frac{4m^2\Delta_{\rm E}(m^2+\Delta_{\rm E}^2)}{(m^2+\Delta_{\rm E}^2)^2+4m^2\Delta_{\rm E}^2}$ & $0$ \T \\[3pt]
$\frac{\psi_1(0^+)}{\psi_2(0^+)}$  &  $\frac{m^2-2m\Delta_0-\Delta_0^2}{m^2+2m\Delta_0-\Delta_0^2}$ & $1$  &  $\big(\frac{m-\Delta_{\rm E}}{m+\Delta_{\rm E}}\big)^2$ & $1$ \T \\[5pt]
\hline
$\frac{\psi_1(0^+)}{\psi_1(0^-)}$  &  $\frac{m^2-2m\Delta_0-\Delta_0^2}{m^2+\Delta_0^2}$  & \; $\frac{m+\Delta_{\rm m}}{m-\Delta_{\rm m}}$ \; & $\frac{m-\Delta_{\rm E}}{m+\Delta_{\rm E}}$ & $\frac{m-i\Delta_{\rm n}}{m+i\Delta_{\rm n}}$ \T \\[3pt]
$\frac{\psi_2(0^+)}{\psi_2(0^-)}$  &  $\frac{m^2+2m\Delta_0-\Delta_0^2}{m^2+\Delta_0^2}$  &  $\frac{m+\Delta_{\rm m}}{m-\Delta_{\rm m}}$ & $\frac{m+\Delta_{\rm E}}{m-\Delta_{\rm E}}$ & $\frac{m-i\Delta_{\rm n}}{m+i\Delta_{\rm n}}$ \T \\[5pt]
\hline\hline
\end{tabular}
\label{table:one}
\end{table}

On the south pole of a TI sphere, a potential $\Delta_0\delta(z)\cdot2v_{\rm z}/m\cdot I$ changes the wave function continuity conditions, integrating Eq.(\ref{eqn:D}) including surface potentials over the vicinity of $z=0$. The amplitudes of $\psi_1$ and $\psi_2$ are either weakened or enhanced across the surface
with changes are always opposite to each other. Consequently, to match the evanescent solutions on the vacuum and the matter sides, the surface band energies are rigidly shifted by $\delta\mathcal{E}_{\rm DP}$. $\tau_{\rm x}$ potentials have similar effects to the $I$-type although $\tau_{\rm x}$ breaks $\mathcal{P}$ symmetry. For a $\tau_{\rm z}$ potential, it simply modifies the mass term on the surface and it is not surprising that it does not affect the surface spectra. A $\tau_{\rm z}$ potential tunes the amplitudes of $\psi_1$ and $\psi_2$ in the same manner on the matter side but this gives no observable effect since $\psi_1(z)=\psi_2(z)$ and $\psi_{1,2}(z<0)\rightarrow 0$ are still valid. Similarly, a potential like $\sigma_{\rm n}\tau_{\rm y}$ ($\Delta_{\rm n}\sigma_{\rm n}\equiv\vec{\Delta}\cdot\vec{\sigma}$) does nothing to the surface spectra and $\psi_1/\psi_2$; however, it couples the two spin flavors and shifts their phase in an opposite way.
Note that $\psi_1=\psi_2$ is always true on the vacuum side since $M\rightarrow\infty$. The above results are summarized in Table \ref{table:one}, providing sufficient information to construct the surface state wave function and to engineer the Dirac point energy position.

On an arbitrary face $\Sigma(\theta)$, the types of surface potentials are the same but their combinations and the corresponding roles are rearranged. This can be fully understood by the fact that the spin-orbital structure ${\bm \tau}\otimes{\bm \sigma}$ on the south pole is replaced by a pseudospin-pseudospin structure ${\bm S}_1(\theta)\otimes {\bm S}_2(\theta)$ on $\Sigma(\theta)$. Note that the combination only occurs for potentials with the same parity.

Although the surface states solutions are stable in the presence of localized surface potentials that preserve $\mathcal{T}$ symmetry, two terms play an essential role in determining the energy position of Dirac point
\begin{eqnarray}
\mathcal{E}_{\rm DP}\!=\!\mathcal{H}^{\rm DP}\!+\!\frac{4m^2(\Delta_0+\Delta_{\rm E})(m^2-\Delta_{0}^2+\Delta_{\rm E}^2)}
{4m^2(\Delta_0+\Delta_{\rm E})^2+(m^2-\Delta_{0}^2+\Delta_{\rm E}^2)^2},
\end{eqnarray}
which implies that $I$ and $\tau_{\rm x}$ potentials (Table \ref{table:one}) are able to tune the Dirac point from the midgap to the band edges $\pm m$ independently.

\indent{\em Discussions.}---
The interactions of the topologically protected bands with surface potentials provides a robust route to engineer and manipulate the topological surface states. In particular, an external surface potential can  raise or lower the Dirac point to the middle of the bulk gap and providing experimental access to the topologically protected band. This goal could be achieved by surface oxidation\cite{Oxidation}, or by other possible chemical processes\cite{Hsan_transport,CO_1,CO_2}
and interactions\cite{self_energy} on the surface. We also point out that the electrostatic gating\cite{CO_2,PJH,Ong_3,Xue_E,Cui}, $E_{\rm eff}\langle r_3\rangle_{\rm sf}$ can act to Stark shift the TI surface state into the gap, with field strength determined by the  penetration length of surface states. Typically, the screened field $\sim 100$ $\rm meV\cdot nm^{-1}$ are required for the surface states evanescent in a couple of quintuple layers.
Our present results provide a framework to study the self-consistent band bending physics of real TI materials\cite{BiSe,BiTe,Cui,Xue,Ando_2} and mean-field models of surface state many-body interactions.

Our model is quite different from the fixed boundary condition (FBC) which arbitrarily clamps the surface state wavefunction to zero at the boundary. The FBC solution is {\em insensitive} to the mass inversion at the surface which topologically protects the surface bands, and consequently it provides no information about the energy of the symmetry protected degeneracy relative to the bulk bands or their interaction with surface-localized potentials.  Furthermore FBC admits an infinite number of (physically spurious) solutions with nonzero energies for $k_1=k_2=0$ that satisfy an (incorrect) surface boundary condition.  By contrast TBC guarantees that there is only one isolated solution, {\em i.e.}, the Dirac point of surface bands protected by the change of bulk topology.  Moreover, TBC demonstrates that the energy position of Dirac point is tunable in the bulk gap via the symmetry allowed scalar terms and surface potentials.

On a non cleavage surface, dangling bonds and their reconstruction may add complexity to the surface state spectrum which can be modeled as surface potentials encoded in the parameters $\Delta_0$ and $\Delta_{\rm E}$. The fingerprint of the novel spin texture near the Dirac point on non cleavage surfaces are determined by the bulk symmetries along with the topological stability of the surface spectrum and may be accessible in ARPES and STM experiments. On an equatorial face, Zeeman coupling or magnetic disorder coupled to the spin degree of freedom do not generally open a gap, and the diamagnetic susceptibility is anticipated to be unusually anisotropic.
More spectacularly, there is intrinsic charge redistribution in the surface bands near the corners of a TI that connect different crystal faces with intrinsically different Dirac point energies.

\indent{\em Acknowledgements.}---This work has been supported by DARPA under grant SPAWAR N66001-11-1-4110.


\begin{thebibliography}{100}
\bibitem{ft1}
 The crossing terms $\propto k_1k_3\sin2\theta$ have negligible effects and thus are ignored for simplicity. Eq.(\ref{eqn:DP}) is an exact result at $\Gamma$ point while $m$ gets reduced by the $m_{1}k_1^2$ and $m_{\shortparallel}k_{\rm y}^2$ terms away from $k=0$.
\end{thebibliography}
\end{document}